\documentstyle[prb,aps,amsfonts,amssymb]{revtex}
\begin{document}
\draft

\title
{
\LARGE \bf
Diffusion and creep of a particle in a random potential}
\author{{\bf D. A. Gorokhov  and  G. Blatter}
\\{\it Theoretische Physik, ETH-H\"onggerberg,
CH-8093 Z\"urich, Switzerland}\\
e-mail: gorokhov@itp.phys.ethz.ch }
\maketitle

\begin{abstract}
{We investigate the diffusive motion of an overdamped classical particle
in a 1D random potential using the mean first-passage time formalism and 
demonstrate the efficiency of this method in the investigation of the 
large-time dynamics of the particle. We determine the $log$-time diffusion 
${\langle{\langle x^{2}(t)\rangle}_{\rm th}\rangle}_{\rm dis}=A\ln^{\beta}
\left ({t}/{t_{r}}\right )$ and relate the prefactor $A,$
the relaxation time $t_{r},$ and the exponent $\beta$
to the details of the (generally non-gaussian) long-range correlated 
potential. Calculating the moments 
${\langle{\langle t^{n}\rangle}_{\rm th}\rangle}_{\rm dis}$
of the first-passage time distribution $P(t),$ we reconstruct the 
large time distribution function itself and draw attention to the phenomenon
of intermittency.
The results can be easily interpreted in terms of the decay of metastable trapped states. In addition, we present a simple derivation
of the mean velocity of a particle moving in a random potential 
in the presence of a constant external force. }
\end{abstract}

{\hskip1.6cm preprint ETH-TH/97-41; accepted for publication in Phys. Rev. B}
\vskip1cm

The motion of an overdamped classical particle in a random
potential 
provides an effective description for
 a variety  of 
phenomena, such as the dynamics of dislocations in solids and of 
domain walls in random field magnets  
(see Ref.\onlinecite{obzor} and references therein),
the relaxation in glasses (see Ref.\onlinecite{Horner} and references
therein), and the electrical transport in
disordered solids (see Ref.\onlinecite{Haus} and references therein).  
Recently, this problem has been considered 
as a phenomenological model
in the context
of glassy dynamics of elastic manifolds in
a quenched random medium\cite{LeDoussal,Scheidl,review}.
In this paper we draw attention to the mean 
first-passage time formalism which appears to be very effective 
for the calculation of different characteristics of the random motion.

We start with the problem of particle 
diffusion in a long-range correlated random potential
and generalize the results for the mean squared diffusion amplitude
$\langle x^{2}(t)\rangle =A\ln^{\beta}\left ({t}/{t_{r}}\right )$
obtained in Refs.\onlinecite{Sinai,Kesten,BCGLD} 
to the case of nongaussian disorder, including the
calculation of the exponent $\beta$ and estimates for the prefactor $A$ and
the relaxation time $t_{r}.$
Second, we investigate the motion 
of a particle subject to a random potential and 
driven by a constant external force.
First, we shall briefly describe the mean first-passage
time formalism.

Consider an overdamped particle in $d$ dimensions subject to the 
potential $V({\bf x})$ and a gaussian random force
$\eta (t)$ with the correlator 
$\langle \eta_{\alpha}(t)\eta_{\beta}(t^{\prime})\rangle =
2T\delta_{\alpha\beta}\delta (t-t^{\prime})$,
$T$ denotes the temperature.
The boundary conditions involve reflecting and absorbing walls
$S_{r}$ and $S_{a},$ respectively. The probability function
$P({\bf x},t)$ is obtained as the solution of the 
Fokker-Planck equation 
\begin{equation}
\frac{\partial P}{\partial t}=
D\frac{\partial^{2} P}{\partial {\bf x}^{2}}
+\frac{\partial}{\partial{\bf x}}
\left (\frac{\partial V}{\partial {\bf x}}P\right ),
\label{Pontr}
\end{equation} 
with
the boundary conditions 
$P({\bf x}\in S_{a})=0$ 
and $\left ({\bf n}\cdot\nabla \right )P({\bf x}\in S_{r})=0,$ 
with ${\bf n}$ the unit vector perpendicular to the surface $S_{r}.$
The diffusion constant $D$ is related to the amplitude
of the random force in the usual way, $D=T$
(we set the particle mobility equal to unity).
Given the initial condition  
$P({\bf x}, t=0)=\delta ({\bf x}-{\bf y}),$ the
$n$-th moment 
$t_{n}({\bf y})={\langle t^{n}({\bf y})\rangle }_{\rm th}$ of the 
 mean absorption time 
$t({\bf y})$
satisfies the Pontryagin equation\cite{Pontryagin,Gardiner,Haenggi}
\begin{equation}
D\frac{\partial^{2}t_{n}}{\partial {\bf y}^{2}}-
\frac{\partial V}{\partial{\bf y}}\frac{\partial t_{n}}{\partial{\bf y}}
=-n t_{n-1}({\bf y}).
\label{PPP}
\end{equation}
The above equation has to  be solved in a closed region with the  
boundary conditions $t_{n}|_{S_{a}}=0$ 
and $\left ({\bf n}\cdot\nabla \right )t_{n}|_{S_{r}}=0,$ see 
Refs.\onlinecite{Pontryagin,Gardiner,Haenggi}.
Eq.~(\ref{PPP}) is equivalent to a chain of equations;
taking into account that $t_{0}=1$ we can determine $t_{1}$
and proceeding by iteration
we find all the moments of 
$t({\bf y})$ 
and, consequently, can reconstruct the probability distribution function for 
$t({\bf y}).$
Let us apply this formalism to the problem of the 1D motion of a particle
in a random environment.

{\it First-passage time  moments and distribution function.}
The equation of motion of a 1D overdamped classical particle
moving in a random potential $U(x)$
takes the form
\begin{equation}
{\dot x}=-\frac{dU}{dx}+\eta (t).
\label{eqofmotion}
\end{equation}
The Fokker-Planck equation 
associated with the stochastic equation (\ref{eqofmotion})
is identical to Eq.~(\ref{Pontr}),
with $D=T,$ 
${ x}\in {\cal R}^{1},$ and $V(x)=U(x).$
Let us consider the case where
$U(x)$ is a gaussian random potential with correlator
${\langle {\left (U(x)-U(y)\right )}^{2}\rangle}_{\rm dis}=K(x-y),$
$K(u)\rightarrow C{|u|}^{\alpha}, $ $u\rightarrow\infty .$
The exponent $\alpha$ is assumed to be positive (if $\alpha <0,$
the disorder merely leads to a renormalization of the 
diffusion coefficient\cite{Haenggi}).
Assume that initially the particle is situated
at the point 
$x=y$ of the interval
$[0,L]$, with 
the boundaries $0$ and $L$ reflecting and $L$ is absorbing
the particle, respectively.
Our aim is to calculate the disorder-averaged mean first-passage
time.  The discrete version of the above problem 
has been studied in a number of 
papers\cite{N1,Murthy,PLD,N2,N3}.
for the case of a random force with $\alpha =1$.

For the case $d=1$ Eq.~(\ref{PPP}) can be solved exactly,
yielding (we remind the reader that $y$ denotes the 
starting point of the particle's diffusion trajectory)
\begin{equation}
t_{n}(y)=\frac{n}{T}\int\limits_{y}^{L}
dy_{1}e^{{U(y_{1})}/{T}}
\int\limits_{0}^{y_{1}}e^{-{U(x_{0})}/{T}}t_{n-1}(x_{0})\thinspace dx_{0}.
\label{tn}
\end{equation}
Expressing the solution for $t_{n-1}(y)$ in terms of 
$t_{n-2}(y),$ substituting into Eq.~(\ref{tn}), and proceeding
iteratively, we obtain the result
\begin{equation}
t_{n}(y)= \frac{n!}{T^{n}}\int_{y}^{L}dy_{n}\int_{0}^{y_{n}}dx_{n-1}
\dots\int_{x_{1}}^{L}dy_{1}\int_{0}^{y_{1}}dx_{0}
\exp{\left \{ \frac{1}{T}
\left (\sum\limits_{i=1}^{n}U(y_{i})
-\sum\limits_{i=0}^{n-1} U(x_{i}) \right ) \right \}}.
\end{equation}
After averaging over the gaussian disorder we arrive at the expression
for the moment ${\langle t_{n}(0)\rangle }_{\rm dis}$
\begin{eqnarray}
{\langle t_{n}(0)\rangle }_{\rm dis} & = &
\frac{n!}{T^{n}}\int_{0}^{L}dy_{n}\int_{0}^{y_{n}}dx_{n-1}
\dots\int_{x_{1}}^{L}dy_{1}\int_{0}^{y_{1}}dx_{0}
\nonumber\\
& \ &
\exp{\left \{ 
\frac{\sum\limits_{i,j} \left [ K(x_{i}-y_{j})
-K(x_{i}-x_{j})
 - K(y_{i}-y_{j})\right ]
}{2T^{2}} \right \}},
\label{tndis}
\end{eqnarray}
where Eq.~(\ref{tndis}) imposes the $2n$ restrictions
\begin{equation}
0\le x_{i-1}\le y_{i},\ \  i=1...n,
\label{restr1}
\end{equation} 
\begin{equation}
x_{i}\le y_{i}\le L, \ \ i=1...n.
\label{restr2}
\end{equation}
In general, the integral in Eq.~(\ref{tndis}) cannot be calculated
exactly. However, in the large distance $L$ limit
we can determine the integral 
by the method of steepest descents,  thus
describing the large-$t$ diffusion.
It can be easily seen that the 
integrand reaches its maximum  
at the point $(x_{i},y_{i})=(0,L)$ (note that the restrictions
(\ref{restr1}) and (\ref{restr2}) are satisfied).
As $K(0)=0$ we obtain
\begin{equation}
{\langle t_{n}(0)\rangle }_{\rm dis}\sim
\exp{\left (\frac{n^{2}K(L)}{2T^{2}}\right )}.
\end{equation}
The prefactor is determined by the functional
dependence close to the saddle-point. Expanding the expression
in the exponent of (\ref{tndis}) and taking into account that 
$K^{\prime}(0)=0$ we arrive at the final result for the
$n$-th moment\cite{typical} of $t$ 
\begin{equation}
{\langle t_{n}(0)\rangle }_{\rm dis}=
\frac{n!}{T^{n}}
{\left (\frac{2T^{2}}{K^{\prime}(L) n}\right )}^{2n}
\exp{\left (\frac{n^{2}K(L)}{2T^{2}}\right )}.
\label{moments}
\end{equation}
The fact that the maximum of the integral in Eq.~(\ref{tndis})
is realized at the boundary manifests itself through
the appearence of the first rather than  second derivative
of $K(x)$ in Eq.~(\ref{moments}).

Using Eq.~(\ref{moments})
we can reconstruct the tails of the probability 
distribution function for the first-passage time $t\equiv 
t(0).$  
We look for a function of the form 
$P(t )\sim
\exp{\left (-A\ln^{\gamma}{\left (t/{\tilde t}_{0}\right )}\right )}.$
Calculating the moments of 
$P(t )$ 
by steepest descents\cite{Mellin} and comparing with Eq.~(\ref{moments}) 
we find
\begin{equation}
P(t )\sim
\exp{\left (-\frac{T^{2}}{2K(L)}
{\ln^{2}{\left ({t}/{{\tilde t}_{0}}\right )}}\right )},
\label{probability}
\end{equation}
 where ${\tilde t}_{0}(T,L)$ is a microscopic time scale 
(while ${\tilde t}_{0}$ accounts for the (dimensional) prefactor in
Eq.~(\ref{moments}), its full dependence on $L$ and $T$ cannot be 
reconstructed from the large-$t$ asymptotics alone). 
The asymptotic expression (\ref{probability})
is applicable for $t\agt {\tilde t}_{0}$ 
and produces 
the strong intermittency observed in the moments 
${\langle t_{n}\rangle}_{\rm dis}$
(see Eq.~(\ref{moments})).
Note that the conjecture 
${\langle t_{n}(0)\rangle }_{\rm dis}\sim
{\left ({\langle t_{1}(0)\rangle }_{\rm dis}\right )}^{n}$
made in Ref.\onlinecite{PLD} is inconsistent with our findings.

{\it Large-time diffusion.} 
Using the result (\ref{probability}) we
can extract a lot of information concerning the large-$t$ behavior
of the particle. Suppose that at $t=0$ the particle is at the point
$x=0.$ Let us estimate its squared 
average displacement after a time $t.$
The characteristic value of $x(t)$
can be found from the implicit equation
$\left ({T^{2}}/{K(x)}\right )\ln^{2}\left ({t}/{{\tilde t}_{0}}\right )\sim 1$
(see Eq.~(\ref{probability}))
defining the 
characteristic value of $x$ where the probability
distribution function $P(t)$ becomes negligible.
With $K(x)\sim C{|x|}^{\alpha},$ we easily find that
\begin{equation}
\langle x^{2}(t)\rangle\sim
{\left (\frac{T^{2}}{C}\right )}^{{2}/{\alpha}}
\ln^{{4}/{\alpha}}\left ({t}/{t_{r}}\right ),
\label{xsquared}
\end{equation}
where $t_{r}$ is a
macroscopic diffusion or relaxation time.
An estimate for $t_{r}$ is obtained by comparing the characteristic
barrier 
$U_{L}\sim\sqrt{CL^{\alpha}}$
 on scale $L$
with the temperature $T.$ This defines the microscopic diffusion
scale 
\begin{equation}
L_{T}\sim {\left ({{T}^{2}}/{C}\right )}^{{1}/{\alpha}}
\end{equation}
and its associated diffusion time 
\begin{equation}
t_{r}\sim {L_{T}^{2}}/{D}\sim \left ({1}/{T}\right )
{\left ({T^{2}}/{C}\right )}^{{2}/{\alpha}}.
\end{equation}

Strictly speaking, the problem of the diffusion on a semi-axis
which we consider here (the boundary $x=0$ is reflecting)
differs from that of the diffusion on the whole axis. However,
the boundary condition at $x=0$ affects the answer (see Eq.~(\ref{xsquared})) only
by a factor of order unity.

Eq.~(\ref{probability}) can be interpreted in terms of the decay
of metastable trapped states. Suppose that a particle 
leaves a metastable state
via thermal activation
over a barrier of random height. Let the barrier distribution function
be gaussian\cite{negatiw}, 
$P(U)=\left ({1}/{\sqrt{2\pi}}\Delta\right )
\exp{\left (-{U^{2}}/{2\Delta^{2}}\right )}.$ Then 
the probability distribution of lifetimes ${\tilde t}_{0}\exp{\left ({U}/{T}\right )}$
is given by the 
expression (we assume that $t>{\tilde t}_{0},$ see also 
Ref.\onlinecite{negatiw})
\begin{equation}
P(t)=\int\limits_{0}^{+\infty}
\frac{dU}{\sqrt{2\pi}\Delta}\exp{\left (-\frac{U^{2}}{2{\Delta}^{2}}\right )}
\delta\left (t-{\tilde t}_{0}e^{{U}/{T}}\right )=
\frac{T}{\sqrt{2\pi}\Delta t}\thinspace 
\exp{\left \{-\frac{T^{2}}{2{\Delta}^{2}}
\ln^{2}\left ({t}/{{\tilde t}_{0}}\right )\right \}}.
\label{decaymet}
\end{equation}
Eq.~(\ref{decaymet})  exhibits the 
same large-$t$ dependence as Eq.~(\ref{probability}), implying
that the diffusion on the scale $L$ is dominated by
one deep potential well of characteristic depth $K^{{1}/{2}}(L).$
This interesting feature of 1D diffusion allows us
to generalize Eq.~(\ref{xsquared}) for the case of non-gaussian disorder.

Assume that the probability for the 
potential difference
$U(x+L)-U(x)$ to be equal to
$E$ is given by the function
\begin{equation}
P(E)=\frac{\delta}{2^{1+{1}/{\delta}}K^{{1}/{\delta}}(L)\Gamma({1}/{\delta})}
\exp{\left \{-\frac{{|E|}^{\delta}}{2K(L)}\right \} },
\label{non-gaussian}
\end{equation} 
with $K(L)\rightarrow CL^{\alpha}, L\rightarrow\infty .$
Calculating $P(t)$ in the same way as in Eq.~(\ref{decaymet}) above,
we arrive at the result (see also Ref.\onlinecite{negatiw})
\begin{equation}
P(t)\sim
\exp{\left \{-\frac{{T}^{\delta}{|\ln
\left ({t}/{{\tilde t}_{0}}\right )
|}^{\delta}}{2K(L)}\right \} },
\label{probability1}
\end{equation} 
implying a $\log t$-diffusion of the form
\begin{equation}
\langle x^{2}(t)\rangle\sim
{\left (\frac{T^{\delta}}{C}\right )}^{{2}/{\alpha}}
\ln^{{2\delta}/{\alpha}}\left ({t}/{t_{r}}\right ),
\label{xsquared1}
\end{equation}
where $t_{r}\sim\left ({1}/{T}\right )
{\left ({T^{\delta}}/{C}\right )}^{{2}/{\alpha}}$
is the relaxation time for non-gaussian disorder.
One can easily verify that Eq.~(\ref{probability1}) implies that
$\ln{\langle t_{n} \rangle }_{\rm dis}\sim n^{{\delta}/{(\delta -1})}.$
For $\delta\rightarrow\infty ,$ the barriers in the system vanish
and the phenomenon of intermittency disappears.

Eq.~(\ref{xsquared1}) is consistent with the exactly solvable  
gaussian random force problem, for which 
the amplitude before the logarithm $A=({61}/{45})\thinspace {T^{4}}/{C^{2}}$
is known exactly
and the relaxation 
time $t_{r}\sim{T^{3}}/{C^{2}},$ see Refs.\onlinecite{obzor,Sinai,Kesten}. 
For the case of gaussian disorder 
($\delta =2$) the exponent ${4}/{\alpha}$ of the logarithm in
 Eq.~(\ref{xsquared1}) has been obtained using RG-techiques\cite{BCGLD}. 
The mean first-passage time method allows us to 
estimate the amplitudes and characteristic relaxation times
and to generalize the results to the case of non-gaussian disorder.
Furthermore,
Eq.~(\ref{probability1}) generalizes the results of previous
investigations of the gaussian random force model\cite{N3}
to the case of arbitrary disorder.

{\it Creep under the action of an external force.}
Let us apply the mean first-passage time formalism
to the problem of creep in 1D.
In a recent paper\cite{LeDoussal}, Le Doussal and Vinokur reported results on
the mobility of an
overdamped classical particle moving in a 1D random potential
$U(x)$ in the presence of a constant external force $f$
(see also Ref.\onlinecite{Scheidl}).
The mean velocity
$V$ has been calculated as a function of $f$ and the correlator
$K(x)$
 of the
random potential. The above model has been considered as  a phenomenological model of glassy dynamics: It turns out that long-range correlations of the random potential lead to the glassy response 
$V\sim\exp{\left (-{1}/{Tf^{\mu}}\right )}$ as $f\rightarrow 0,$
with $T$ the temperature and $\mu >0$ a constant, whereas in the case of short-range correlations $V\sim f$ as $f\rightarrow 0.$

The problem has been solved\cite{LeDoussal} using a 
generalization of the method
introduced by Derrida\cite{Derrida} for discrete models,
assuming that the random potential
is a periodic function of the coordinate $x,$
$U(x)=U(x+L).$ Next, the stationary solution 
${\tilde P}(x)$
of the Fokker-Planck equation has been found
for a fixed current ${\tilde J}.$ 
Using the conditions ${\tilde P}(0)={\tilde P}(L),$
$U(0)=U(L),$ $V={\tilde J}L,$ 
the normalization condition on ${\tilde P}(x),$
and taking the limit $L\rightarrow\infty$, the authors 
of Ref.\onlinecite{LeDoussal}
arrive at the result (see also Ref.\onlinecite{Scheidl})
\begin{equation}
\frac{1}{V}=
\frac{1}{T}\int\limits_{0}^{\infty}ds e^{-{fs}/{T}}
{\langle
e^{{\left (U(x+s)-U(x)\right )}/{T}}\rangle }_{x},
\label{promezh}
\end{equation}
where 
\begin{equation}
{\langle A\rangle }_{x}=
({1}/{L})\lim_{L\to\infty }\int\limits_{0}^{L}dx A(x).
\end{equation}
 For the case of a 
gaussian random potential with
${\langle {\left (U(x)-U(y)\right )}^{2}\rangle}_{\rm dis}=K(x-y)$ 
the averaging
procedure can be easily performed, yielding the final result
\begin{equation}
\frac{1}{V}=\frac{1}{T}\int\limits_{0}^{\infty}
dx\exp{\left (-\frac{fx}{T}+\frac{K(x)}{2T^{2}}\right )}.
\label{final} 
\end{equation}
Let us show how 
Eqs.~(\ref{promezh}) and (\ref{final}) can be obtained
in a simple and elegant way using the first-passage 
time method. 
This technique has several advantages: {\it i}) it does not rely on
the periodic continuation of the random potential and, consequently,
one does not have to worry 
about the commutation of the two limits $t\rightarrow\infty$ and
$L\rightarrow\infty $; {\it ii}) instead of solving the
stationary Fokker-Planck equation with a fixed current one can use
the well-known solutions of the 1D Pontryagin
equation\cite{Pontryagin,Gardiner,Haenggi} and simply average them over
the disorder; {\it iii}) using the first-passage time technique
one can investigate the finite size effects, the moments of the 
average velocity
distributon function
etc., i.e., the information obtained is much richer.

Returning to the 1D problem of an overdamped particle subject
to the potential $V(x)=U(x)-fx$ we can solve Eq.~(\ref{PPP})
for the first moment exactly\cite{Pontryagin,Gardiner,Haenggi},
\begin{equation}
t_{1}(y)=\frac{1}{T}\int\limits_{y}^{L}dz\thinspace e^{{V(z)}/{T}}
\int\limits_{a}^{z}dx\thinspace e^{-{V(x)}/{T}}.
\label{time}
\end{equation}
Here we assume that the point $L$ is absorbing, the point $a$ is 
reflecting\cite{comment}, and $a<y<L.$ In the limit $a\rightarrow -\infty$
the particle does not feel the left boundary 
as the external force $f$ is chosen to be positive.
Averaging Eq.~(\ref{Pontr})
over the disorder we obtain
\begin{equation}
{\langle t_{1}(y)\rangle }_{dis}=\frac{1}{T}
\int\limits_{y}^{L}dz \thinspace e^{-{fz}/{T}}
\int\limits_{-\infty}^{z}dx\thinspace  e^{{fx}/{T}}
{\langle e^{{\left (U(z)-U(x)\right )}/{T}}\rangle }_{x}.
\label{timetime}
\end{equation}
If the random potential distribution is spacially
homogeneous, the average ${\langle e^{{\left [U(z)-U(x)\right ]}/{T}}\rangle }_{\rm dis}$
is a function of $|z-x|$
and is the same as the translation average 
${\langle e^{{\left [U(x+s)-U(x)\right ]}/{T}}\rangle }_{x}$
(see Eq.~(\ref{promezh})),
where we have  
introduced the new variable $s=x-z.$  
Eq.~(\ref{timetime}) then takes the form
\begin{equation}
{\langle t_{1}(y)\rangle }_{\rm dis}
=\frac{L-y}{T}\int\limits_{0}^{\infty}ds\thinspace
e^{{-fs}/{T}}
{\langle e^{{\left (U(x+s)-U(x)\right )}/{T}}\rangle }_{\rm dis}
.
\label{tdis}
\end{equation}
If the integral in Eq.~(\ref{tdis}) converges, the 
mean first passage time averaged over disorder
is an extensive quantity.  In the   
limit ${L-y}\rightarrow\infty$
we can apply the central limit theorem, implying that
${1}/{\langle t_{1}(y)\rangle }_{\rm dis}=
{\langle {1}/{t_{1}(y)}\rangle }_{\rm dis}.$
The mean velocity $V$ is simply 
${\langle {(L-y)}/{t_{1}(y)}\rangle }_{\rm dis}$
and we can easily see that Eq.~(\ref{tdis}) becomes equivalent to 
Eq.~(\ref{promezh}).
For a  gaussian random potential,
\begin{equation}
{\langle t_{1}(y)\rangle }_{\rm dis}=\frac{L-y}{T}
\int\limits_{0}^{\infty}ds\thinspace
\exp{\left (-\frac{fs}{T}+\frac{K(s)}{2T^{2}}\right )},
\end{equation}
which is equivalent to Eq.~(\ref{final}).

{\it Relation to discrete models.}
Let us compare the 
 mean first passage-time  averaged over disorder 
obtained using Eq.~(\ref{time}) with that found for a discrete
1D random walk in a random force-field.
Consider a 1D disordered lattice. Let us denote by $p_{n}$
the probability of hopping forward $n\rightarrow n+1$ and by
$q_{n}=1-p_{n}$ that of hopping back $n\rightarrow n-1.$
The index $n$ enumerates the sites of the lattice.
For the random-force problem\cite{obzor},
\begin{equation}
p_{n}=\frac{\exp{\left [\frac{dF_{n+1}}{2T}\right ]} }
{\exp{\left [-\frac{dF_{n}}{2T}\right ]}+
\exp{\left [\frac{dF_{n+1}}{2T}\right ]} 
}, 
\label{hopping}
\end{equation}
with $d$ the lattice spacing 
(note that in this model {\it both} time and space are discrete).
The force $F$ is a gaussian random variable
satisfying the conditions 
$\langle F_{n}F_{m}\rangle ={\mu \delta_{nm}}/{d}$
and $\langle F_{n}\rangle =0.$ Note that 
$\langle \ln\left ({p_{n}}/{q_{n}}\right )\rangle =0,$ 
see Refs.\onlinecite{obzor,Sinai}. 
The point $n=0$ is supposed to be reflecting
and $N={L}/{d}$ is absorbing.
In Ref.\onlinecite{PLD} the estimate
${\langle t_{1}(0)\rangle }_{\rm dis}\sim\exp{\left [\beta L\right ]},$
where
 $\beta=\ln{\langle {q_{n}}/{p_{n}}\rangle }={\mu}/{4T^{2}}$ and
$L \rightarrow\infty $, has been obtained.

The continuous version of this problem  is 
described by Eq.~(\ref{eqofmotion}) with a
correlator for the random potential given by $K(x-y)=\mu |x-y|.$
Averaging Eq.~(\ref{time}) over disorder we obtain  
\begin{equation}
{\langle t_{1}(y)\rangle }_{\rm dis}  = 
\frac{4T^{3}}{\mu^{2}}
\left [\exp{\left (\frac{\mu L}{2 T^{2}}\right )}-
\exp{\left (\frac{\mu y}{2 T^{2}}\right )}\right ]\nonumber\\
  - \frac{2 T}{\mu}\left (L-y\right ).
\end{equation}
For $y$ close to the reflecting boundary $0$
and for large $L$ the result takes
the simple form
\begin{equation}
{\langle t^{(1)}(y)\rangle }_{\rm dis}\rightarrow
\frac{4T^{3}}{\mu^{2}}\exp{\left (\frac{\mu 
L
}{2T^{2}}\right )}.
\label{asymptotic}
\end{equation}
We find that for both (discrete in space and time and continuous) cases 
$\ln {\langle t_{1}(0)\rangle }_{\rm dis}\sim L.$
The coefficient of proportionality, however, is different, 
$\beta={\mu}/{2T^{2}}$ for the continous model versus
$\beta={\mu}/{4T^{2}}$ for the discrete case.

We thus have arrived at a different asymptotic behavior
for the average first-passage time.
The origin of this difference is found in the inequivalence
of the discrete in space and time and the discrete in space--continuous
time models: For the latter, consider 
the lattice master equation describing the probability $P_{n}(t)$
for the particle to be on the site $n$:
\begin{equation}
\frac{dP_{n}}{dt} = 
W_{n,n+1}P_{n+1}+W_{n,n-1}P_{n-1}
 - 
\left (W_{n+1,n}+W_{n-1,n}\right )P_{n},
\label{master}
\end{equation}
where the hopping probabilities $W_{n,n\pm 1}$ and $W_{n\pm 1, n}$
are determined by the potential\cite{obzor} (as the lattice spacing tends to
zero, one easily recovers the continuous
 Fokker-Planck equation (\ref{Pontr}) and thus this model
is equivalent to the continuous in space and time model).
The terms $W_{n,n+1}P_{n+1}$ and $W_{n,n-1}P_{n-1}$
in Eq.~(\ref{master})
describe the hopping $n+1\rightarrow n$ and $n-1\rightarrow n,$
respectively.
The contribution
  $\left (W_{n+1,n}+W_{n-1,n}\right )P_{n}$ 
describes the probability to stay at the same site $n.$
In a next step let us also discretize the time variable $t$. 
Suppose that at time $t,$ $P_{n}=\delta_{nm}.$
In the limit of a finite but small time step $\Delta t$ the probabilities
for the processes. $m\rightarrow m\pm 1$ 
behave as $\alpha_{1}\Delta t$ and $\alpha_{2}\Delta t$ 
($\alpha_{1}$ and $\alpha_{2}$ are two constants),
hence the probability for a particle to stay at the site $m$ after
$t\rightarrow t+\Delta t$ is $1-(\alpha_{1}+\alpha_{2})\Delta t,$ 
i.e. the particle
most likely stays at the same site.
On the other hand, for the discrete lattice walks considered in 
Refs.\onlinecite{N1,PLD,N2,N3} the probability of staying at the same site vanishes
by definition: The particle may only jump to the neighbouring sites.
Thus the  continuous random walk cannot be obtained as a
limiting case of a discrete random walk when {\it both}
the time and the lattice are discrete.

Briefly summarizing, we have investigated the motion of an overdamped
classical
particle in a random potential. The moments of the first-passage time 
averaged over gaussian disorder
have been found and 
are given by Eq.~(\ref{moments}).
The asymptotic form of the first-passage time
distribution function has been reconstructed, see Eq.~(\ref{probability}),
 allowing us to find the large-$t$ dependence of $\langle x^{2}(t)\rangle $,
see Eq.~(\ref{xsquared}). The results obtained can be 
easily interpreted in terms of the decay of metastable states.
The latter feature has allowed us to generalize the results for   
 $\langle x^{2}(t)\rangle $ to the case of non-gaussian disorder,
see Eq.~(\ref{xsquared1}). In addition, we 
have presented a simple derivation
of the mean velocity of a particle driven by a constant external force
across a disordered medium
and have discussed
the relation to the discrete time
random walk on a lattice.

We thank Stefan Scheidl and Valerii Vinokur for discussions.


\begin{thebibliography}{99}
\bibitem{obzor}J.~P.~Bouchaud, A.~Comtet, A.~Georges, and P.~Le~Doussal,
Ann. Phys. {\bf 201}, 285 (1990).
\bibitem{Horner}H.~Horner, Z. Phys. B {\bf 100}, 243 (1996).
\bibitem{Haus}
S.~Alexander, J.~Bernasconi, W.~R.~Schneider, and R.~Orbach, Rev. Mod. Phys.
{\bf 53}, 175 (1981);
J.~W.~Haus and K.~W.~Kehr, Phys. Rep. {\bf 150}, 263 (1987).
\bibitem{LeDoussal}P.~Le~Doussal and V.~M.~Vinokur, Physica C {\bf 254},
63 (1995).
\bibitem{Scheidl}S.~Scheidl, Z. Phys. B {\bf 97}, 345 (1995).
\bibitem{review}for a review see G. Blatter, M.V. Feigel'man, V.B. Geshkenbein,
A.I. Larkin, and V.M. Vinokur, Rev. Mod. Phys. {\bf 66}, 1125 (1994).
\bibitem{Sinai}Ya.~G.~Sinai, Theory Probab. Appl. {\bf 27}, 247 (1982).
\bibitem{Kesten}H.~Kesten, Physica A {\bf 138}, 299 (1986).
\bibitem{BCGLD}
J.~P.~Bouchaud, A.~Comtet, A.~Georges, and P.~Le~Doussal,
J. Phys. {\bf 48}, 1445 (1987). 
\bibitem{Pontryagin}L.~Pontryagin, A.~Andronov, and A.~Vitt,
Zh. Eksp. Teor. Fiz. {\bf 3}, 165 (1933). 
\bibitem{Gardiner}C.~W~Gardiner, {\it Springer Series in Synergetics, Vol.~13:
Handbook of Stochastic Methods for Physics, Chemistry, and Natural Science,}
Springer-Verlag, Berlin (1983).
\bibitem{Haenggi}P.~H\"anggi, P.~Talkner, and B.~Borkovec, Rev. Mod.
Phys. {\bf 62}, 251 (1990).
\bibitem{N1}S.~H.~Noskowicz and I. Goldhirsch, 
Phys. Rev. Lett. {\bf 61}, 500 (1988).
\bibitem{Murthy}K.~P.~N.~Murthy and K.~W.~Kehr, Phys. Rev. A {\bf 40},
2082 (1989).
\bibitem{PLD}P.~Le~Doussal, Phys. Rev. Lett. {\bf 62}, 3097 (1989).
\bibitem{N2}S.~H.~Noskowicz and I. Goldhirsch, 
Phys. Rev. Lett. {\bf 62}, 3098 (1989).
\bibitem{N3}S.~H.~Noskowicz and I. Goldhirsch, 
Phys. Rev. A {\bf 42}, 2047 (1990).
\bibitem{typical}We wish to point out that the first moment
${\langle t_{1}\rangle}\sim\exp{\left ({K(L)}/{2T^{2}}\right )}$
determines the {\it average} first-passage time, while the 
${\it typical}$ time $t_{\rm typ}$
of the first-passage which controls the large
time diffusion scales as $\ln t_{\rm typ}\sim{\sqrt{K(L)}}/{T}$,
see also Refs.\onlinecite{N1,PLD}.
\bibitem{Mellin}The function $P(t)$
should satisfy the conditions: {\it i}) $P(t)$ is continuous;
{\it ii}) tends to zero sufficiently fast such that all its moments exist
for arbitrary $n$. The reconstruction of the probability
distribution function $P(t)$ involves
the inverse Mellin transformation:
Let $\langle t_{n}\rangle =\int_{0}^{\infty}t^{n}P(t)dt$
be a regular function of the variable $n=\sigma + i\tau$
in the strip $\sigma_{1}<\sigma<\sigma_{2}$ and 
$\int_{-\infty}^{+\infty}|\langle t_{\sigma +i\tau}\rangle |d\tau <\infty$,
then $P(t)=\left ({1}/{2\pi i}\right )
\int_{c-i\infty}^{c+i\infty}t^{-n-1}\langle t_{n}\rangle\thinspace  dn$,
with $\sigma_{1}<c<\sigma_{2}$.
For the function 
$\langle t_{n}\rangle$ given by Eq.~(\ref{moments})
the main term in the asymptotics
for $P(t\rightarrow\infty$) is determined by the exponential 
$\exp{\left ({n^{2}K(L)}/{2T^{2}}\right )}$.
The integral in the inversion formula then can be calculated by 
steepest descents.
\bibitem{negatiw}The 
distribution function $P(U)$ for the random barrier height $U$
(see also Eq.~(\ref{non-gaussian}))
allows for negative values of $U$ describing
the absence of a barrier. However, in the limit of  strong 
metastability (large variation of the random barrier 
$\Delta$ or small temperatures $T$) this feature of
$P(U)$  has a negligible effect on $P(t),$ see Eqs.~(\ref{decaymet})
and (\ref{probability1}) below.  
\bibitem{Derrida}B.~Derrida, J. Stat. Phys. {\bf 31}, 433 (1983). 
\bibitem{comment}In the limit
$a\rightarrow -\infty$ the left point is irrelevant,
i.e., it can be either reflecting or absorbing. 

\end{thebibliography}
\end{document}